# Optical spin orientation and depolarization in Ge

C. Rinaldi,[*] M. Cantoni, and R. Bertacco

*LNESS-Dipartimento di Fisica, Politecnico di Milano, Via Anzani 42, 22100 Como, Italy*

**Abstract**

Optical spin orientation and depolarization phenomena in semiconductors are of overwhelming importance for the development of spin-optoelectronics. In this paper we employ Ge-based spin-photodiodes to investigate the room temperature spectral dependence of optical spin orientation in Germanium, in the range 400-1550 nm, and the photo-carrier spin relaxation phenomena. Apart from the maximum in the spin polarization of photo-carriers for photon energy resonant with the direct gap (1550 nm), we experimentally demonstrate the presence of a second sizable peak at 530 nm due to photo-generation far away from the center of the Brillouin zone, within the L valleys. Furthermore, our data indicate that the equivalent the spin lifetime of holes in Ge is in the order of 5-10 ps, meaning that the spin diffusion length for holes at room temperature is larger than expected, in the order of 150-220 nm.

[*]christian.rinaldi@mail.polimi.it

Spin-optoelectronics is a novel research area aiming to add a new degree of freedom, the photon helicity, to optoelectronics. By exploiting the interplay between the photon angular momentum and the spin of electrons in spintronic devices, integrated emitters (spin-LEDs) and detectors (spin-photodiodes) of circularly polarized light have been proposed. In these devices the control and detection of the polarization state are achieved via the control of the magnetization of ferromagnetic electrodes, without use of external optical elements. The combination of these elements in novel architectures opens new frontiers to integrated communication technology applications.[1] Recently, a GaAs-based optical communication system has been demonstrated, where circularly polarized light is emitted by a spin-LED and detected by a spin-photodiode (spin-PD)[2]. Despite GaAs is traditionally the direct gap favorite material for spin-optoelectronics, recently Ge has attracted a considerable attention, mainly because of its weaker spin-orbit coupling leading to longer spin coherence time.[3] Moreover, spin manipulation,[4] spin transport,[5] spin optical pumping in the infrared [6,7,8,9,10,11] and electrical spin injection[12] in Ge and SiGe heterostructures have been reported. In our previous works[13,14] we reported the first demonstration of the room temperature operation of a spin-PD based on the fully epitaxial Fe/MgO/Ge(001) heterostructure, working at 1300 nm wavelength.[15,16] The coupling between the photon helicity and the spin of carriers optically injected in Ge ("optical spin orientation")[17] enables to convert the information related to the optical polarization into the spin polarization of photo-generated carriers. Then the Fe/MgO/Ge tunneling junction acts as efficient spin-filter of carriers, since the tunneling transmission depends on the direction of their spin with respect to the Fe magnetization. The degree of circular polarization thus results to be proportional to the photocurrent variation for opposite orientations of the Fe magnetization. A percentage variation of the photocurrent on the order of 6% has been measured at room temperature for a light wavelength of 1300 nm, both in forward and reverse bias. Different physical mechanism are involved in the spin-PD operation: (i) Magnetic Circular Dichroism, (ii) spin-optical orientation, (iii) depolarization due to spin-flip events, (iv) spin-dependent tunneling across the barrier. While (i) and (iv) have been widely studied, both experimentally and theoretically,[6,7,8,9,10,11], a comprehensive study of the interplay between optical spin orientation and depolarization has not been reported so far. Noteworthy, there is still a poor understanding of the electrons and holes optical spin orientation in Ge, especially at photon energies much higher than the direct gap (0.8 eV), where the optical spin injection can lead to electrons populating not only the Γ but also the satellite L valleys. The optical spin orientation in Ge has been theoretically investigated by Rioux and Sipe,[18] who calculated the degree of spin orientation achievable by pumping the semiconductor with circularly polarized light on a wide spectral range. However, no experimental data concerning optical spin orientation far from the $\Gamma$ point of the Brillouin zone have been reported so far for Ge. Even for GaAs, only very recently the photon energy dependence well above the absorption edge (around the $L$ point of the Brillouin zone) has been reported at 10 K.[19]

In this Letter, we discuss the physics of the optical spin orientation and depolarization in Ge, via measurements on Ge-based spin-PDs. We investigated, at room temperature, the spectral dependence of optical spin orientation in a wide photon energy range (0.8 -3.1 eV). Surprisingly enough, the highest percentage variation of the photocurrent in our spin-PDs upon reversal of the light circular polarization is far away from the absorption onset, where the highest initial spin polarization could be obtained. We measured a maximum variation of about 10% at 530 nm (~2.3 eV) which is related to the optical pumping in the L valley of the band structure (Fig. 1a), where Rioux and Sipe predict a local maximum of the optical spin orientation.[18] This unexpected result comes from a compromise between the needs of (i) a high degree of spin polarization of photo-generated carriers and (ii) a reduced spin depolarization during transport towards the MgO barrier. Ideally, photon absorption should take place in a thin layer close to the barrier, much thinner than the spin diffusion length, and this can be obtained at high photons energies, corresponding to small light attenuation lengths. The fitting of our data with a diffusive model including a photo-generation term allows to confirm the theoretical prediction[18] of a maximum in the spin-optical orientation at 2.3 eV, as well to confirm the relatively high value of the equivalent spin diffusion length for holes in Ge at 300K (150-220 nm).[13]

The Fe/MgO/Ge heterostructures have been prepared by Molecular Beam Epitaxy, as discussed in details elsewhere.[15,20] We employed lightly n-doped Ge substrates (the resistivity was ~47 Ω·cm) to maximize the spin diffusion length. Spin-PDs with circular shape and different areas have been fabricated via optical lithography and ion beam etching.

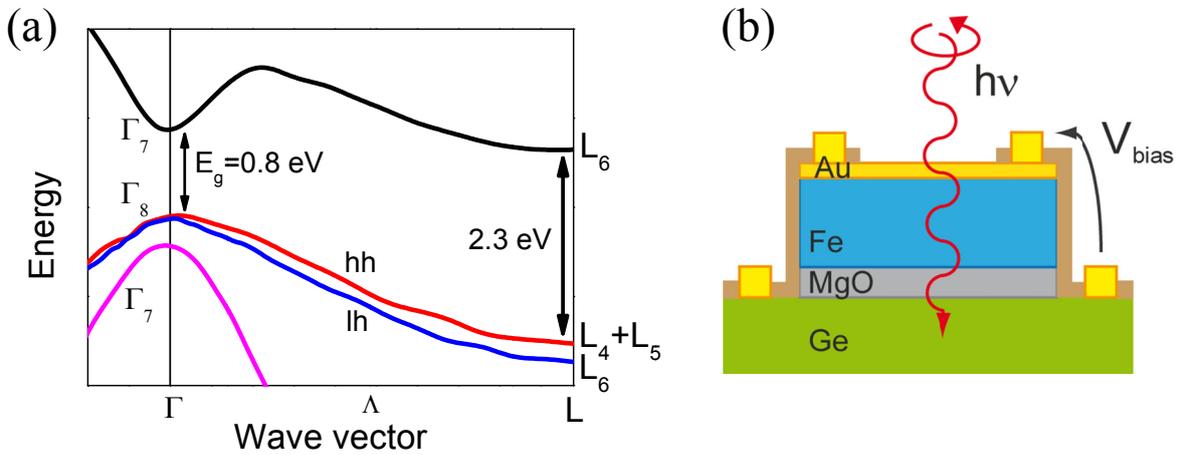

**Fig. 1.** (a) Band structure of Ge from the Γ to the L point of the Brillouin zone (along the direction Λ)[21]. (b) Sketch of a Fe/MgO/Ge spin-photodiode.

A sketch of a spin-photodiode is shown in Fig. 1b.[13] A laser beam with different wavelength, from 400 nm to 1550 nm, impinges perpendicularly on the photodiodes surface. The helicity is modulated between the left (σ−) and right (σ+) circular polarization with a photoelastic modulator operating at a frequency of 50 kHz. The helicity-dependent photocurrent flowing across the photodiode ($\Delta I = I^{\sigma+} - I^{\sigma-}$) is measured using a lock-in amplifier, while the magnetization of the Fe layer is driven out of plane, parallel or anti-parallel, to the photon angular momentum. $\Delta I$ can be expressed as:[13]

$$\Delta I(V_{bias}) = 2\sigma \cdot I_{photo} \cdot (D + A_{SF}) \tag{1}$$

where $I_{photo}(V_{bias})$ is the photocurrent measured for incident linearly polarized light and in-plane magnetization, $D$ is the Magnetic Circular Dichroism (MCD) asymmetry due to dichroic absorption from the Fe layer, $A_{SF}$ the spin dependent transport asymmetry.

In Fig. 2a we present the experimental values of $\Delta I$ as a function of the applied bias ($V_{bias}$) in case of photodiodes with MgO thickness of 1.8 nm illuminated with 543 nm light. For positive (negative) bias the electric field in the depletion region drives photo-generated electrons (holes) towards the MgO barriers. This implies that the photocurrent modulation arises from the spin filtering of electrons (holes) in forward (reverse) bias. The spin transport asymmetry $A_{SF}$ is plotted in Fig. 2b after subtraction of the MCD contribution (see Supplementary Information). As already discussed in Ref. 13 for 1300 nm excitation wavelength, also at 543 nm we observe a sizable spin filtering both for holes (5.5±0.5% at -0.4 V) and electrons (3.3±0.5% at +0.4 V) thus indicating similar values of their spin diffusion lengths.

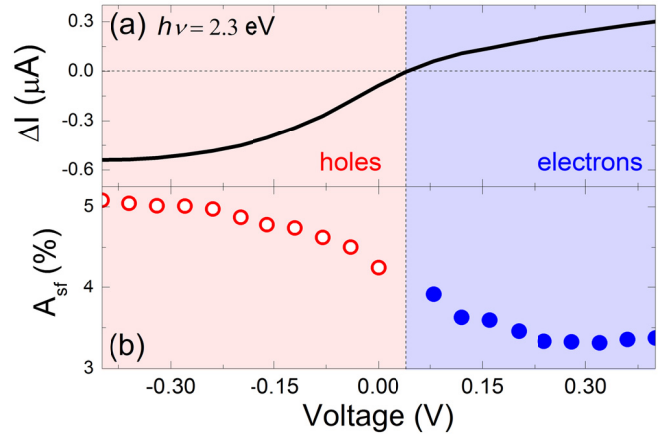

**Fig. 2. (a)** Photocurrent variation $\Delta I$ as a function of the bias voltage due to the full reversal of the circular polarization of light in the case of $\lambda$= 543 nm light ($h\nu$= 2.3 eV). **(b)** Spin-filtering asymmetry ($A_{sf}$) deduced from the data of panel (a) and after the subtraction of the MCD contribution.

In Fig. 3, the $A_{SF}$ values measured in the spectral region from $h\nu= 0.8$ eV (1550 nm) to $h\nu = 3.1$ eV (400 nm), both for holes ($V_{bias}$= -0.4 V, red empty dots) and for electrons ($V_{bias}$= +0.4 V, blue filled dots) are reported. Noteworthy $A_{SF}$ is slightly higher for holes than for electrons in all the investigated photon energy range. This indicates that our previous suggestion of a sizable equivalent spin diffusion length for holes is an intrinsic result, independent on the particular photon energy used for spin optical pumping. Even more interestingly, data in Fig.3a show a non-monotone evolution of $A_{SF}$ vs. photon energy, which sheds light on the physical process of spin optical pumping. Surprisingly enough, $A_{SF}$ presents a relative maximum at about 1 eV, slightly above the absorption edge of Ge (~0.8 eV), and then the absolute maximum at ~2.3 eV, in the visible range, before assuming small negative values above 2.4 eV. This is quite unexpected, because the maximum degree of optical spin orientation ($P_S$) is located at the direct band gap of the semiconductor according to $k \cdot p$ calculations[18]. In order to shed light on the apparent contradiction between the calculated degree of optical spin orientation and the measured values of $A_{SF}$, we developed a diffusive model of spin-PDs including spin-optical pumping.[13] Even though this represents only a first order approximation for a semiconductor such as Ge, it contains the essential physics to describe the spectral response of our devices. In the supplementary information we use this model to work out a suitable expression for $A_{SF}$, taking into account all the various physical phenomena involved: spin optical pumping, propagation of photo-generated carriers, depolarization and finally tunneling across the barrier. The spin filtering asymmetry is given by

$$A_{SF} = 2 P_S \frac{l_{sf}^{SC}}{l_{sf}^{SC} + \lambda_L} \frac{\gamma r_B r_{SC}}{R_{SC}(r_{SC} + r_B) + r_B(1-\gamma^2) r_B^2 + r_B r_{SC}} \quad (2)$$

where $P_S$ is the carriers spin polarization immediately after the photo-generation;[18] $\lambda_L$ is the light absorption length;[22] $\gamma$ is the spin-dependent interfacial resistance asymmetry; $r_B$ is the resistance per unit surface of the tunneling barrier; $R_{sc}=\rho_{sc}L$ is the product between the resistivity of the semiconductor ($\rho_{sc}$) and the length of the semiconducting side of the device ($L$); $r_{sc}=\rho_{sc}l_{sf}$ is the product between $\rho_{sc}$ and the spin diffusion length ($l_{sf}$) in the semiconductor of the carriers involved in tunneling. $l_{sf}$ is defined as $l_{sf} = \sqrt{D\tau_{sf}}$, where $D$ is the diffusion coefficient and $\tau_{sf}$ the spin-flip time of the carriers.

While the role of barrier related parameters ($\gamma$ and $r_B$) has been extensively investigated[13], in this Letter we focus on the photon energy dependence of $A_{SF}$ via two parameters: (i) the initial degree of spin-polarization $P_S(h\nu)$ by optical orientation and (ii) the ratio between the light absorption length $\lambda_L(h\nu)$ and the spin diffusion length $l_{sf}$. From Eq. 2 it is possible to distinguish two operating regimes for spin-PDs, leading to simplified expressions for $A_{SF}$:

$$A_{SF} \sim \begin{cases} P_S \dfrac{l_{sf}}{\lambda_L} & \text{where } \lambda_L \gg l_{sf} \\ P_S & \text{where } \lambda_L \ll l_{sf} \end{cases} \qquad (3)$$

where the typical value of $l_{sf}$ to be considered are 1 μm for electrons and 150-220 nm for holes in the case of our slightly doped n-Ge substrate[13]. It is evident that, apart from the linear dependence on $P_S$, $A_{SF}$ depends on the ratio between $\lambda_L$ and $l_{sf}$. For $\lambda_L \gg l_{sf}$, in the regime that we call *spin depolarization regime*, carriers are photo-generated in a deep layer and suffer from a major depolarization during their motion towards the barrier, so that $A_{SF}$ is strongly suppressed. On the contrary, if $\lambda_L \ll l_{sf}$, in the true *optical spin orientation regime*, carriers are photo-generated in a layer much thinner than the spin diffusion length, so that they preserve their initial spin polarization till they reach the tunneling barrier. As $\lambda_L$ strongly depends on the photon energy, the latter has a big impact on the photo-generated carrier depolarization. In Fig. 4, $\lambda_L{}^{22}$ and $P_S{}^{18}$ are plotted as a function of the photon energy. In the same figure the spin diffusion length for holes and electrons are marked with continuous lines. The spin depolarization regime, where $\lambda_L \gg l_{sf}$, corresponds to photon energies close to the band gap of the semiconductor. Here the joint density of states for optical transitions is low, resulting in a value of $\lambda_L$ (~12.5 μm at 0.8 eV[22]) very large with respect to the typical $l_{sf}$ values (1 μm for electrons and 150-220 nm for holes). $\lambda_L$ then rapidly decreases when the photon energy increases. As a result, $A_{SF}$ decreases when moving from 1 eV to 0.8 eV, despite the degree of spin polarization increases towards the absorption edge.

Above the absorption edge, $P_S$ first decreases and then reaches a secondary local maximum around $h\nu = 2.3$ eV. In fact, this energy nearly corresponds to transitions at the $L$ point of the Brillouin zone, where the crystal field reduces the symmetry in such a way that a splitting between heavy holes and light holes takes place, so that in a relatively narrow spectral region it is possible to achieve very high spin-polarization within the $L$ valleys. Unfortunately, with such a photon energy the generation of carriers is not restricted only to the $L$ valleys but involves several points of the Brillouin zone, so that the net degree of spin polarization reduces to $P_S \sim 20\%$ (according to Ref. 18). Moreover, at 2.3 eV the absorption length is reduced by two orders of magnitude with respect to the case of 0.8 eV: in fact, there is a large region in the $k$ space where the $L_6$ conduction and $L_{4,5}$ valence bands are parallel, with effective masses larger than those at the Γ point. The absorption at this energy is very effective so that $\lambda_L \ll l_{sf}$ and we definitely enter the spin-orientation regime. With this in mind, we can understand why at 2.3 eV we observe the absolute maximum of $A_{SF}$: the lower $P_S$ (20%) with respect to the value at 0.8 eV (50%) is largely compensated by the much lower depolarization of photo-generated carriers during the propagation towards the barrier. The fit of our data with the diffusive model is also shown in Fig. 3 with dashed line for forward bias (electrons) and continuous line for reverse bias (holes). The fit parameters for electrons are $\tau_{sf}$= 120 ps, $\gamma$= 0.3 and $D$= 0.01035 m$^2$s$^{-1}$, while for holes are: $\tau_{sf}$ =5 ps, $\gamma$=0.9 and $D$= 0.0049 m$^2$s$^{-1}$.[13] The values of the other parameters employed for the calculation are reported the Supplementary Information.

The fit is in a very nice agreement with the experiment in the case of holes (Fig. 3), thus confirming our previous findings[13] that the equivalent spin relaxation time for this type of carrier is definitely found to be larger than previously expected, in the order of 5 ps. Noteworthy such a long equivalent spin lifetime for holes is in agreement with recent non-local measurements[23] and sheds light on this controverted question. Moreover, we can exclude that this long lifetime is related to localized interfacial states rather than free holes,[11] because our model that gives a good fit of $A_{SF}(\lambda)$ relies on the hypothesis that the carriers are distributed in the semiconductor according to the absorption length of light. We can notice that at $h\nu$= 2.3 eV the absolute maximum of $A_{SF}$ found is lower with respect to the value of the fit. We explain this in terms of intervalley relaxation: at these energies, holes are primarily generated at the L point of the Brillouin zone and then relax to the valence band maximum located at Γ, where the transport takes place; since the intervalley scattering is a momentum scattering process assisted by phonons, we expect a partial relaxation of the spins.[6]

The fit of $A_{SF}$ for electrons (Fig. 3) is instead in a fair agreement with experimental data. Contrary to the case of holes, the fit can reproduce very well the maximum at 2.3 eV but the expected values in the infrared part of the spectrum (spin depolarization regime) are always higher than the experimental ones. The situation is specular to that of holes: in such energy range, electrons are photo-generated in Γ relax towards L and the photon-assisted momentum scattering between the two point is detrimental for the spin, preserving only a fraction of the initial degree of spin polarization produced by optical orientation. The electron spin-phonon interaction has been only recently addressed theoretically and not yet experimentally. As pointed out by J.-M. Tang et al.[6], the intervalley scattering at room temperature represents the limiting factor for the spin lifetime of carriers in Ge. Experimental work on this topic need to be done to confirm calculations. Finally, we have to underline that our model is based on the hypothesis of a diffusive regime, while in a spin-PD the drift caused by the electric field of the depletion region can eventually play a role.

Even more interestingly, our data represent the first experimental proof of theoretical predictions for the spectral dependence of $P_S$ in Ge on the whole spectral range. The enhancement of $P_S$ at the L point of the Brillouin zone, however, is characteristic of every semiconductor with the diamond/zinc-blend crystal structure (e.g., Si, Ge, GaAs, GaP, CdTe),[18,24] making this operating regime very interesting for spin-optoelectronics.

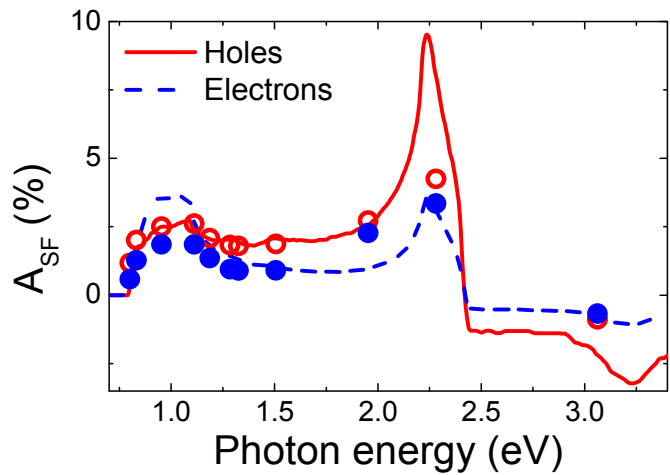

**Fig. 3** The spin filtering asymmetry of Ge-based spin-PD's is reported versus photon energy for holes (open red dots) and electrons (closed blue dots). A fit of $A_{sf}$ with the extended Fert-Jaffres model is also provided for both the carriers (solid red line and dashed blue line, respectively).

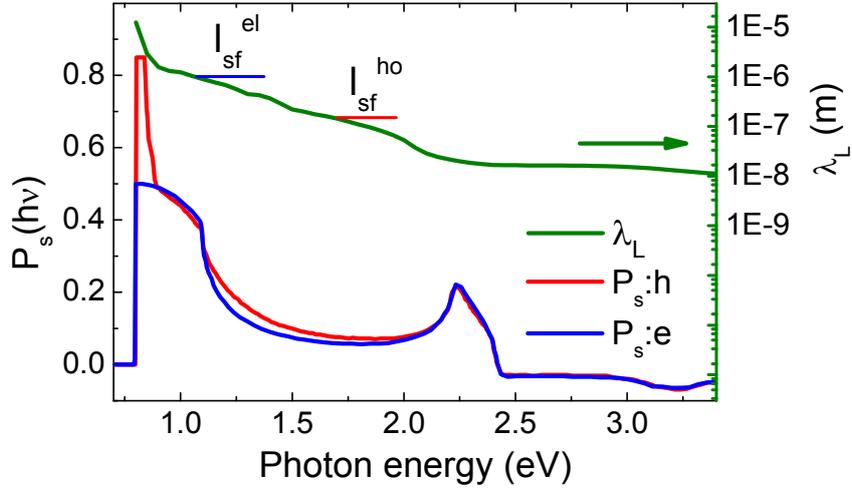

**Fig. 4.** Degree of optical spin orientation $P_S$ for holes and electrons (red and blu solid lines, respectively) as calculated by Rioux and Sipe[18]. The light attenuation length $\lambda_L$22 (green solid line) is reported and compared with the spin-flip length $l_{sf}$, in order to identify the two working regime for both the carriers.

In conclusion, we demonstrated that a ferromagnet/oxide/semiconductor heterostructure represent a valid tool for a deep study of the optical spin orientation and depolarization. We experimentally proved that, apart from the expected high value of the degree of spin polarization by optical pumping for a photon energy close to the gap of Ge, there is another maximum at 2.3 eV in the visible range. Furthermore our data taken in a wide spectral range definitely demonstrate that the equivalent spin relaxation time for holes in Ge is much longer than expected, on the order of 5 ps. Finally this work demonstrates that Ge-based spin PDs can be used as suitable integrated detectors of the photon helicity in a wide spectral range spanning from the infrared (1550 nm) to the visible (530 nm).

The authors thank G. Isella, S. Sanvito, N. M. Caffrey and D. Petti for fruitful discussions, M. Leone for his skillful technical support, and M. Marangoni, C. Manzoni, G. Cerullo and A. Melloni for help with the optical measurements. This work was partially funded by the project FIRB OSSIDI NANOSTRUTTURATI: MULTI-FUNZIONALITA' E APPLICAZIONI (RBAP115AYN).

# Supplementary information

**S.1: Subtraction of the Magnetic Circular Dichroism (MCD) from the helicity-dependent photocurrent**

The MCD contribution can be properly subtracted in order to obtain information concerning the spin filtering of the spin-photodiode. In particular, $A_{SF}=\Delta I(V)/(2I_{photo})-D$ is the percentage variation of the photocurrent due to spin-filtering of carriers upon full reversal of the light helicity from right to left. The coefficient $D$ for $\lambda$= 1300 nm ($h\nu$= 0.95 eV) has been directly measured from the dichroic adsorption of a 10 nm-thick Fe sample, grown by MBE on a MgO substrate. For every wavelength we employed the tabulated values of $D(\lambda)$ from the literature,[25] normalized to the actual MCD measured at 1300 nm for our Fe films. The resulting MCD versus photon energy is shown in Fig. S1. We want to underline that the trend of $A_{SF}(h\nu)$ is significantly different from that of the dichroism. Moreover, the MCD contribution is smaller than $A_{SF}$ for every photon energy We can then exclude that the trend of $A_{SF}$ reported in Fig. 3 could be due to the a wrong evaluation of the dichroic coefficient $D$.

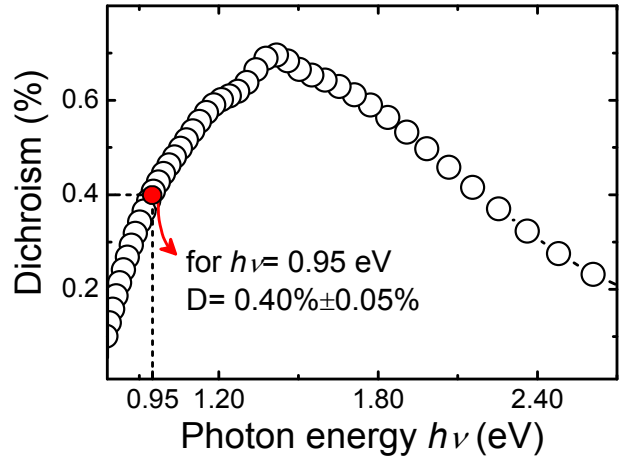

**Fig. S1**. Dichroism of an epitaxial film of Fe(10 nm) grown on a MgO substrate. The tabulated data from literature have been renormalized to fit the measurements performed with $h\nu$= 0.95 eV ($\lambda$=1300 nm) and a saturating field $H$= 2T (the red point indicated in the graph).

## S.2: Calculation of the spin-dependent transport asymmetry ($A_{SF}$) from the modified Fert-Jaffrès model

In our previous work[13] we developed an extended Fert-Jaffrès model[26] in order to treat the effect of spin optical pumping, propagation of photo-generated carriers, depolarization and finally tunneling across the barrier in any spin-PD based on a FM/B/SC heterostructure, where FM, B and SC are the ferromagnet, the insulating barrier and the semiconductor, respectively.

In Eq. 14 of Ref. 13, Supporting Information, we calculated the variation in the photocurrent at fixed bias and magnetization upon reversal of the light circular polarization:

$$\Delta J^{\sigma} = (J^{\sigma=+1} - J^{\sigma=-1}) = \frac{2e(\eta_{\sigma}^{-} - \eta_{\sigma}^{+})\frac{r_{SC} l_{sf}^{SC}}{\alpha_L l_{sf}^{SC} + 1} \cdot \frac{r_B}{r_{SC} + r_B}}{R_{SC} + r_B(1-\beta\gamma) + r_B \frac{\beta r_{SC} + (\beta - \gamma) r_B}{r_{SC} + r_B}} \quad (1S)$$

$\eta_{\sigma}^{+(-)}$ are the spin-dependent photo-generation rates per unit volume depending on the polarization state ($\sigma$); $(\eta_{\sigma}^{+} - \eta_{\sigma}^{-})/(\eta_{\sigma}^{+} + \eta_{\sigma}^{-}) = -(+)P_S$ for $\sigma$=+(-)1, respectively, where $P_S$ is the photo-carrier polarization immediately after photo-generation. $\alpha_L = 1/\lambda_L$ is the light absorption coefficient. $\bar{r}_{SC}$ is the product between the resistivity of the semiconductor and the length of the semiconducting side of the device; in the text is indicated as $R_{sc}=\rho_{sc}L$. $\beta$ is the spin dependent resistance asymmetry of the ferromagnet. Other parameters are described in the text.

Expression (1S) can be re-written, with different notation, as

$$\Delta J^{\sigma} = \frac{2P_S e(\eta_{\sigma}^{-} + \eta_{\sigma}^{+})\lambda_L \frac{l_{sf}^{SC}}{l_{sf}^{SC} + \lambda_L} r_B r_{SC}}{\bar{r}_{SC}(r_B + r_{SC}) + r_B(1-\beta\gamma)(r_B + r_{SC}) + r_B(\beta r_{SC} + (\beta - \gamma) r_B)} \quad (2S)$$

The photocurrent density is given by

$$J_{photo} = e(\eta_{\sigma}^{-} + \eta_{\sigma}^{+})\lambda_L \quad (3S)$$

so that the spin-dependent transport asymmetry is

$$A_{SF} = 2P_S \frac{l_{sf}^{SC}}{l_{sf}^{SC} + \lambda_L} \frac{r_B r_{SC}}{R_{SC}(r_{SC} + r_B) + r_B(1-\gamma^2)r_B^2 + r_B r_{SC}} \quad (4S)$$

that is Eq. 2 in the text.

# References

[1] M. Oestreich et al., "Spintronics: Spin Electronics and Optoelectronics in Semiconductors", *Advances in Solid State Physics* **41**, 173-186 (2001)

[2] R. Farshchi, M. Ramsteiner, J. Herfort, A. Tahraoui, and H. T. Grahn, "Optical communication of spin information between light emitting diodes", *Appl. Phys. Lett.* **98**, 162508 (2011), doi:10.1063/1.3582917

[3] Gordon, J. P., and Bowers, K. D., "Microwave Spin Echoes from Donor Electrons in Silicon", *Phys. Rev. Lett.* **1**, 368 (1958).

[4] Ganichev, S. D., Danilov, S. N., Bel'kov, V. V., Giglberger, S., Tarasenko, S. A., Ivchenko, E. L., Weiss, D., Jantsch, W., Schäffler, F., Gruber, D., and Prettl, W., "Pure spin currents induced by spin-dependent scattering processes in SiGe quantum well structures", *Phys. Rev. B* **75**, 155317 (2007).

[5] Shen, C., Trypiniotis, T., Lee, K. Y., Holmes, S. N., Mansell, R., Husain, M., Shah, V., Li, X. V., Kurebayashi, H., Farrer, I., de Groot, C. H., Leadley, D. R., Bell, G., Parker, E. H. C., Whall, T., Ritchie, D. A., and Barnes, C. H. W., "Spin transport in germanium at room temperature", *Appl. Phys. Lett.* **97**, 162104 (2010).

[6] Tang, J.-M., Collins, B. T., and Flatté, M. E. "Electron spin-phonon interaction symmetries and tunable spin relaxation in silicon and germanium", *Phys. Rev. B* **85**, 045202 (2012).

[7] Loren, E. J., Ruzicka, B. A., Werake, L. K., Zhao, H., van Driel, H. M., and Smirl, A. L., "Optical injection and detection of ballistic pure spin currents in Ge", *Appl. Phys. Lett.* **95**, 092107 (2009).

[8] Loren, E. J., Rioux, J., Lange, C., Sipe, J. E., van Driel, H. M., and Smirl, A. L., "Hole spin relaxation and intervalley electron scattering in germanium", *Phys. Rev. B* **84**, 214307 (2011).

[9] Bottegoni, F., Isella, G., Cecchi, S., and Ciccacci, F., "Spin polarized photoemission from strained Ge epilayers", *Appl. Phys. Lett.* **98**, 242107 (2011).

[10] Bottegoni, F., Ferrari, A., Isella, G., Cecchi, S., Marcon, M., Chrastina, D., Trezzi, G., and Ciccacci, F., "Ge/SiGe heterostructures as emitters of polarized electrons", *J. Appl. Phys.* **111**, 063916 (2012).

[11] F. Pezzoli et al., *Phys. Rev. Lett.* **108**, 156603 (2012).

[12] Liu, E.-S., Nah, J., Varahramyan, K. M., and Tutuc, E., "Lateral Spin Injection in Germanium Nanowires", *Nano Lett.* **10**, 3297-3301 (2010).

[13] C. Rinaldi, M. Cantoni, D. Petti, A. Sottocorno, M. Leone, N. M. Caffrey, S. Sanvito and R. Bertacco, *Adv. Mat.* **24**, 3037 (2012).

[14] C. Rinaldi, M. Cantoni, D. Petti, and R. Bertacco, *J. Appl. Phys.* **111**, 07C312 (2012).

[15] D. Petti *et al.*, *J. Appl. Phys.* **109**, 084909 (2011).

[16] M. Cantoni *et al.*, *Appl. Phys. Lett.* **98**, 032104 (2011).

[17] M. I. Dyakonov and V. I. Perel in *Optical Orientation*, edited by F. Meier and B. P. Zakharchenya (North-Holland, Amsterdam 1984).

[18] J. Rioux and J. E. Sipe, *Phys. Rev. B* **81**, 155215 (2010).

[19] T. Zhang et al., L-valley Electron Spin Dynamics in GaAs. *ArXiv*:1207.5978v1.

[20] M. Cantoni *et al.*, *Microel. Eng.* **88**, 530 (2011).

[21] W. Elder and R. Ward, J. Zhang. *Phys. Rev. B* **83**, 165210(2011).

[22] Dash, W. C., and Newman, R., "Intrinsic optical absorption in Single-Crystal Germanium and Silicon at 77K and 300 K", *Phys. Rev.* **99**, 1151 (1955).

[23] H. Saito, S. Watanabe, Y. Mineno, S. Sharma, R. Jansen, S. Yuasa, and K. Ando. "Electrical creation of spin accumulation in p-type germanium". *Solid State Commun.* **151**, 1159 (2011).

[24] F. Nastos et al., *Phys. Rev. B* **76**, 205113 (2007).

[25] G. S. Krinchik, V. A. Artem'ev, *Sov. Phys. JETP* **26**, 1080 (1968).

[26] A. Fert and H. Jaffrès, *Phys. Rev. B* **64**, 184420 (2001) .